\documentclass[aps,preprint]{revtex4}%
\usepackage{amsfonts}
\usepackage{amsmath}
\usepackage{amssymb}
\usepackage{graphicx}%
\begin{document}
\title{{\LARGE ON \ THE ZERO-ENERGY\ UNIVERSE }}
\author{Marcelo Samuel Berman$^{1}$}
\affiliation{$^{1}$Instituto Albert Einstein/Latinamerica\ - Av. Candido Hartmann, 575 -
\ \# 17}
\affiliation{80730-440 - Curitiba - PR - Brazil - email: msberman@institutoalberteinstein.org}
\date{20 June 2009}

\begin{abstract}
We consider the energy of the Universe, from the pseudo-tensor point of
view(Berman,1981). We find zero values, when the calculations are
well-done.The doubts concerning this subject are clarified, with the novel
idea that the justification for the calculation lies in the association of the
equivalence principle, with the nature of co-motional observers, as demanded
in Cosmology. In Section 4, we give a novel calculation for the zero-total
energy result.

\end{abstract}
\maketitle

\begin{center}

{\LARGE ON \ THE ZERO-ENERGY UNIVERSE }

\bigskip\bigskip Marcelo Samuel Berman
\end{center}

\bigskip

{\LARGE 1. Introduction}

In\bigskip\ pages 90 and 91 of \ the best-seller (Hawking, 2001), Hawking
describes inflation (Guth, 1981), as an accelerated expansion of the Universe,
immediately after the creation instant,while the Universe, as it
expands,borrows energy from the \ gravitational field to create more matter.
According to his \ description, the positive matter energy is exactly balanced
by the negative gravitational energy, so that the total energy is zero,and
that when \ the size of the Universe doubles, both the matter and
gravitational energies also double, keeping the total energy zero (twice
zero). \textbf{Our task will be to show why the Universe is a
zero-total-energy entity, by means of pseudo-tensors.}

\bigskip

\bigskip\bigskip The pioneer works of Nathan Rosen (Rosen, 1994), Cooperstock
and Israelit, (1995) , showing that the energy of the Universe is zero, by
means of calculations involving pseudotensors, and Killing vectors,
respectively, are here given a more simple approach. We shall show that the
energy of the Robertson-Walker's Universe is zero, (Berman, 2007). Berman
(1981) worked as a pioneer, in pseudotensor calculations for the energy of
Robertson-Walker's Universe. He made the calculations on which the present
paper rest, and, explicitly obtained the zero-total energy for a closed
Universe, by means of LL-pseudotensor.

\bigskip

The zero-total-energy of the Roberston-Walker's Universe, and of any Machian
ones, have been shown by many authors (Berman 2006; 2006a; 2007; 2007a;
2007b). \bigskip It may be that the Universe might have originated from a
vacuum quantum fluctuation. In support of this view, we shall show that the
pseudotensor theory (Adler et al, 1975) points out to a null-energy for a
Robertson-Walker-flat Universe, in a Cartesian-coordinates calculation.
(Berman, 2006; 2006a; 2007; 2007a; 2007b; Rosen, 1995; York Jr, 1980;
Cooperstock, 1994; Cooperstock and Israelit, 1995; Garecki,1995; Johri et
al.,1995; Feng and Duan,1996; Banerjee and Sen,1997; Radinschi,1999;
Cooperstock and Faraoni,2003). Next, we shall show that in spherical
coordinates, we would obtain a wrong result, but see also Katz (2006, 1985);
Katz and Ori (1990); and Katz et al (1997). Recent developments include
torsion models (So and Vargas, 2006), and, a paper by Xulu(2000).

The reason for the failure of curvilinear coordinate energy calculations
through pseudotensor, resides in that curvilinear coordinates carry non-null
Christoffel symbols, even in Minkowski spacetime, thus introducing inertial or
fictitious fields that are interpreted falsely as gravitational
energy-carrying (false) fields.

Carmeli et al.(1990) listed four arguments against the use of Einstein%
\'{}%
s pseudotensor:1.the energy integral defines only an affine vector;2.no
angular-momentum is available;3. as it depends only on the metric tensor and
its first derivatives, it vanishes locally in a geodesic system;4. due to the
existence of a superpotential, which is related to the total conserved
pseudo-quadrimomentum, by means of a divergence, then the values of \ the
metric tensor, and its first derivatives, only matter, on a surface around the
volume of the mass-system.

\bigskip We shall argue below that, for the Universe, local and global Physics
blend together. The pseudo-momentum, is to be taken like the linear momentum
vector of Special Relativity, i.e., as an affine vector. If the Universe \ has
some kind of rotation, the energy-momentum calculation refers to a co-rotating
observer. Such being the case, we go ahead for the actual calculations.

\bigskip

{\LARGE 2. Energy of the flat Robertson-Walker's Universe}

\bigskip

\bigskip

\bigskip It has been generally accepted that the Universe has zero-total
energy. The first such claim, as far as the present author recollects, was due
to Feynman(1962-3). Lately, Berman(2006, 2006 a) has proved this result by
means of simple arguments involving Robertson-Walker's metric for any value of
the tri-curvature ( $0,-1,1$ ).

The pseudotensor \ $t_{\nu}^{\mu}$\ , also called Einstein's pseudotensor, is
such that, when summed with the energy-tensor of matter \ $T_{\nu}^{\mu}$\ \ ,
gives the following conservation law:\ \ \ 

\bigskip

$\left[  \sqrt{-g}\left(  T_{\nu}^{\mu}+t_{\nu}^{\mu}\right)  \right]  ,_{\mu
}=0$ \ \ \ \ \ \ \ \ \ \ \ \ \ \ \ \ \ \ \ \ .\ \ \ \ \ \ \ \ \ \ \ \ \ \ \ \ \ \ \ \ \ \ \ \ \ \ \ \ \ \ \ \ \ \ \ \ \ \ \ \ \ \ \ \ \ \ \ (1)

\bigskip

In such case, the quantity

\bigskip

$P_{\mu}=\int\left\{  \sqrt{-g}\left[  T_{\mu}^{0}+t_{\mu}^{0}\right]
\right\}  $ $d^{3}x$ \ \ \ \ \ \ \ \ \ \ \ \ \ \ \ \ \ \ \ \ ,\ \ \ \ \ \ \ \ \ \ \ \ \ \ \ \ \ \ \ \ \ \ \ \ \ \ \ \ \ \ \ \ \ \ \ \ \ \ \ \ \ (2)

\bigskip

is called the general-relativistic generalization of the energy-momentum
four-vector of special relativity (Adler et al, 1975).

\bigskip

It can be proved that \ $P_{\mu}$\ \ \ is conserved when:

\bigskip

a) \ $T_{\nu}^{\mu}\neq0$\ \ only in a finite part of space; \ \ \ \ \ \ and,\ \ 

b) \ $g_{\mu\nu}\rightarrow\eta_{\mu\nu}$ \ when we approach infinity, where
\ $\eta_{\mu\nu}$\ \ is the Minkowski metric tensor.

\bigskip

However, there is no reason to doubt that, even if the above conditions were
not fulfilled, we might eventually get a constant \ $P_{\mu}$\ , because the
above conditions are sufficient, but not strictly necessary. We hint on the
plausibility of other conditions, instead of a) and b) above.

\bigskip

Such a case will occur, for instance, when we have the integral in (2)\ \ is
equal to zero.

\bigskip

For R.W.'s flat metric, we get exactly this result, because, from Freud's
(1939) formulae, we have

\bigskip

$P_{\nu}=\frac{1}{2\kappa}\int\sqrt{-g}\{$ $[$ $\delta_{\nu}^{0}\left(
g^{\beta\alpha}\Gamma_{\beta\rho}^{\rho}-g^{\beta\rho}\Gamma_{\beta\rho
}^{\alpha}\right)  +\delta_{\nu}^{\alpha}\left(  g^{\beta\rho}\Gamma
_{\rho\beta}^{0}-g^{0\rho}\Gamma_{\rho\beta}^{\beta}\right)  -$

$\bigskip$

$\ \ \ \ \ \ \ \ \ \ \ \ \ \ \ \ \ \ \ \ \ \ \ \ \ \ \ \ \ -\ \left(
g^{\beta\alpha}\Gamma_{\beta\nu}^{0}-g^{\beta0}\Gamma_{\beta\nu}^{\alpha
}\right)  ]\}$ $_{^{\prime}\alpha}d^{3}x$\ \ \ \ \ \ \ \ \ \ \ . \ \ \ \ \ \ \ \ \ \ \ \ \ \ \ \ \ \ \ (3)\ \ \ \ 

\bigskip

\bigskip From R.W.'s flat metric,

\bigskip

$ds^{2}=dt^{2}-R^{2}(t)d\sigma^{2}$ \ \ \ \ \ \ \ \ \ \ \ \ \ \ \ \ \ \ \ \ \ \ ,\ \ \ \ \ \ \ \ \ \ \ \ \ \ \ \ \ \ \ \ \ \ \ \ \ \ \ \ \ \ \ \ \ \ \ \ \ \ \ \ \ \ \ \ \ \ \ (4)

\bigskip

we find that, with no index summation,

\bigskip

$g^{ii}\Gamma_{ii}^{0}\equiv g^{00}\Gamma_{0i}^{i}$
\ \ \ \ \ \ \ \ \ \ \ \ \ \ \ \ \ \ \ \ \ \ \ \ \ \ \ , \ \ \ \ \ \ \ \ \ \ \ \ \ \ \ \ \ \ \ \ \ \ \ \ \ \ \ \ \ \ \ \ \ \ \ \ \ \ \ \ \ \ \ \ \ \ \ \ \ \ (5)

\bigskip

and, then,

\bigskip

$P_{i}=0$ \ \ \ \ \ \ \ \ \ \ \ \ \ \ \ \ \ \ \ ( $i=1,2,3$ \ ) \ \ \ \ \ \ \ \ \ \ \ \ \ \ \ \ .\ \ \ \ \ \ \ \ \ \ \ \ \ \ \ \ \ \ \ \ \ \ \ \ \ \ \ \ \ \ \ \ \ \ (6)

\bigskip

On the other hand, considering only the non-vanishing Christoffel symbols, we
would find, taken care of (4),

$P_{0}=0$ \ \ \ \ \ \ \ \ \ \ \ \ \ \ \ . \ \ \ \ \ \ \ \ \ \ \ \ \ \ \ \ \ \ \ \ \ \ \ \ \ \ \ \ \ \ \ \ \ \ \ \ \ \ \ \ \ \ \ \ \ \ \ \ \ \ \ \ \ \ \ \ \ \ \ \ \ \ \ \ \ \ \ \ \ \ \ \ \ \ (7)

\bigskip

Because we found a constant result, we may say that the total energy of a flat
R.W.'s Universe is null.

\bigskip

A different calculation, as follows, leads to the same result. Weinberg(1972) defines:

\bigskip

$h_{\mu\nu}\equiv g_{\mu\nu}-\eta_{\mu\nu}$
\ \ \ \ \ \ \ \ \ \ \ \ \ \ \ \ \ \ \ , \ \ \ \ \ \ \ \ \ \ \ \ \ \ \ \ \ \ \ \ \ \ \ \ \ \ \ \ \ \ \ \ \ \ \ \ \ \ \ \ \ \ \ \ \ \ \ \ \ \ \ \ \ \ \ \ \ \ (8)

\bigskip

and then solves for the 4-pseudo-momentum, obtaining:

\bigskip

$P^{j}=-\frac{1}{16\pi G}\int\left\{  -\frac{\partial h_{kk}}{\partial
t}\delta^{ij}+\frac{\partial h_{k0}}{\partial x^{k}}\delta_{ij}-\frac{\partial
h_{j0}}{\partial x^{i}}+\frac{\partial h_{ij}}{\partial t}\right\}  \left\{
n_{i}r^{2}d\Omega\right\}  $ \ \ \ \ \ \ \ \ \ \ \ \ , \ \ \ \ \ \ \ \ (9)

\bigskip

and,

\bigskip

$P^{0}=-\frac{1}{16\pi G}\int\left\{  \frac{\partial h_{jj}}{\partial x^{i}%
}-\frac{\partial h_{ij}}{\partial x^{j}}\right\}  \left\{  n_{i}r^{2}%
d\Omega\right\}  $ \ \ \ \ \ \ \ \ \ \ \ \ \ \ \ , \ \ \ \ \ \ \ \ \ \ \ \ \ \ \ \ \ \ \ \ \ \ \ \ \ \ \ \ \ \ \ (10)

\bigskip

with

\bigskip

$d\Omega=\sin\theta d\theta d\phi$\ \ \ \ \ \ \ \ \ \ \ , \ \ \ \ \ \ \ \ \ \ \ \ \ \ \ \ \ \ \ \ \ \ \ \ \ \ \ \ \ \ \ \ \ \ \ \ \ \ \ \ \ \ \ \ \ \ \ \ \ \ \ \ \ \ \ \ \ \ \ \ \ \ \ \ \ \ (11)

\bigskip

and, \ 

\bigskip

$n_{i}\equiv\frac{X_{i}}{r}$\ \ \ \ \ \ \ \ \ \ \ \ \ \ \ \ \ \ .\ \ \ \ \ \ \ \ \ \ \ \ \ \ \ \ \ \ \ \ \ \ \ \ \ \ \ \ \ \ \ \ \ \ \ \ \ \ \ \ \ \ \ \ \ \ \ \ \ \ \ \ \ \ \ \ \ \ \ \ \ \ \ \ \ \ \ \ \ \ \ \ (12)\ \ \ 

\bigskip

Though (9) \ and (10) can be constants in the case considered in Weinberg's
book, it is evident that if the integrals in both (9) and (10) are null, we
still can call the null result of (10) as a proof of the null energy of the
R.W. flat Universe. And, in this case,

\bigskip

$P^{i}=P^{0}=0$ \ \ \ \ \ \ \ \ \ \ \ \ ( \ $i=1,2,3$\ \ )
\ \ \ \ \ \ \ \ \ \ \ . \ \ \ \ \ \ \ \ \ \ \ \ \ \ \ \ \ \ \ \ \ \ \ \ \ \ \ \ \ \ \ \ \ \ \ \ \ \ (13)

\bigskip

A similar result would be obtained from Landau-Lifshitz pseudotensor (1975),
where we have:

\bigskip

$P_{LL}^{\nu}=\int(-g)\left[  T^{\nu0}+t_{L}^{\nu0}\right]  $ $\ d^{3}x$
\ \ \ \ \ \ \ \ \ \ \ \ \ \ \ \ , \ \ \ \ \ \ \ \ \ \ \ \ \ \ \ \ \ \ \ \ \ \ \ \ \ \ \ \ \ \ \ \ \ \ \ \ \ \ \ \ \ \ \ \ \ (14)

\bigskip

where,

\bigskip

$(-g)t_{L}^{ik}=\frac{1}{2\kappa}\{$ $\ g_{^{\prime}l}^{ik}$ $g_{^{\prime}%
m}^{lm}-g_{^{\prime}l}^{il}$ $g_{^{\prime}m}^{km}+\frac{1}{2}g^{ik}%
g_{lm}g^{ln\text{ }},_{\rho}g_{^{\prime}n}^{\rho m}-(g^{il}g_{mn}g_{^{\prime
}\rho}^{kn}g_{^{\prime}l}^{m\rho}+g^{kl}g_{mn}g_{^{\prime}\rho}^{in}%
g_{^{\prime}l}^{m\rho})+$

\bigskip

\ \ \ \ \ \ \ \ \ \ \ \ \ \ \ \ \ \ \ \ \ $+$ $g_{lm}g^{n\rho}g_{^{\prime}%
n}^{il}g_{^{\prime}p}^{km}+\frac{1}{8}(2g^{il}g^{km}-g^{ik}g^{lm})(2g_{n\rho
}g_{qr}-g_{\rho q}g_{nr})g_{^{\prime}l}^{nr}g_{^{\prime}m}^{pq}$ \ $\}$ \ \ \ \ ,

\bigskip

(in this last expression all indices run from \ $0$ \ \ to $\ \ 3$)
\ \ \ \ \ \ \ \ . \ \ \ \ \ \ \ \ \ \ \ \ \ \ \ \ \ (15)

\bigskip

A short calculation shows that:

\bigskip

$P_{LL}^{\nu}=0$ \ \ \ \ \ \ \ \ \ \ \ \ \ \ \ \ ( $\nu=0,1,2,3$%
\ \ )\ \ \ \ \ \ \ \ . \ \ \ \ \ \ \ \ \ \ \ \ \ \ \ \ \ \ \ \ \ \ \ \ \ \ \ \ \ \ \ \ \ \ \ \ \ \ (16)\ \ \ \ \ \ \ \ \ 

\bigskip

The above results could also follow from superpotential formulae (Freud,
1939). For instance, from the Einstein's superpotential:

\bigskip

$P_{\nu}=\int\left[  U_{\nu}^{[0\sigma]}\right]  _{,\text{ }\sigma}d^{3}x$ \ \ \ \ \ \ \ \ \ \ \ \ \ \ \ ,

\bigskip

where,

\bigskip

$2\kappa\sqrt{-g}\underset{(E)}{U_{\nu}^{[\mu\rho]}}=g_{\nu\sigma}\left\{
g\left[  g^{\mu\sigma}g^{\rho\lambda}-g^{\mu\lambda}g^{\rho\sigma}\right]
\right\}  _{,\lambda}$ \ \ \ \ \ \ \ \ \ \ \ \ .

\bigskip

Then, we find, for the Robertson-Walker's metric,

\bigskip

$\underset{(E)}{U_{\nu}^{[0\sigma]}}=0$ \ \ \ \ \ \ \ \ \ \ \ \ \ \ \ (
\ $\nu=0,1,2,3$\ \ \ ) \ \ .

\bigskip

Then, \ $P_{0}=0$\ \ \ .\ Analogously, we would find \ $P_{i}=0$\ \ \ \ .\ \ 

\bigskip

\bigskip

\bigskip{\LARGE 3. Counter-examples in Energy Calculations}

\bigskip

\textbf{3A. Closed Robertson-Walker's Counter-Example:}

\bigskip

\bigskip We can give a counter-example, showing that if we do not use
Cartesian coordinates, but other system, say, spherical coordinates, the
energy calculation becomes flawed (Berman, 1981), as it has been warned by
Weinberg(1972) and Adler, Bazin and Schiffer(1975), among others.

\bigskip

Consider a closed Robertson-Walker's metric:

\bigskip

$ds^{2}=-\frac{R^{2}(t)}{\left(  1+\frac{r^{2}}{4}\right)  ^{2}}\left[
dr^{2}+r^{2}d\theta^{2}+r^{2}\sin^{2}\theta d\phi^{2}\right]  +dt^{2}$ \ \ \ . \ \ \ \ \ \ \ \ \ \ \ \ \ \ \ (17)

\bigskip

With the energy momentum tensor for a perfect fluid, whose comoving components are:

\bigskip

$T_{0}^{0}=\rho$ \ \ \ \ \ \ ,

\bigskip

$T_{1}^{1}=T_{2}^{2}=T_{3}^{3}=-p$ \ \ \ \ \ \ ,

\bigskip

$T_{\nu}^{\mu}=0$ \ \ \ if \ \ \ \ $\mu\neq\nu$ \ \ \ \ \ ,

\bigskip

\bigskip where \ \ $\rho$\ \ \ and \ $p$\ \ stand for energy density and
cosmic pressure, respectively, and with a pseudo-tensor given by:

\bigskip

$\sqrt{-g}t_{\beta}^{\alpha}=\frac{1}{2\kappa}\left[  \delta_{\beta}^{\alpha
}U-g_{^{\prime}\beta}^{\mu\nu}\frac{\partial U}{\partial g_{^{\prime}\alpha
}^{\mu\nu}}\right]  $ \ \ \ \ \ \ , \ \ \ \ \ \ \ \ \ \ \ \ \ \ \ \ \ \ \ \ \ \ \ \ \ \ \ \ \ \ \ \ \ \ \ \ \ \ \ \ \ (18)

\bigskip

where,

\bigskip

$U=\sqrt{-g}g^{\mu\nu}\left[  \Gamma_{\mu\alpha}^{\beta}\Gamma_{\nu\beta
}^{\alpha}-\Gamma_{\mu\nu}^{\alpha}\Gamma_{\alpha\beta}^{\beta}\right]  $ \ \ \ \ \ ,\ \ \ \ \ \ \ \ \ \ \ \ \ \ \ \ \ \ \ \ \ \ \ \ \ \ \ \ \ \ \ \ \ \ \ \ \ \ \ \ (19)

\bigskip

we shall find a time-varying result for the energy.

\bigskip

If we consider Einstein's field equations, with \ $k=+1$\ \ , where
\ $k$\ \ is the tricurvature, in particular we have:

\bigskip

$3H^{2}=\kappa\rho-3R^{-2}$\ \ \ \ \ \ \ \ \ \ \ \ \ \ \ \ \ ,\ \ \ \ \ \ \ \ \ \ \ \ \ \ \ \ \ \ \ \ \ \ \ \ \ \ \ \ \ \ \ \ \ \ \ \ \ \ \ \ \ \ \ \ \ \ \ \ \ (20)

\bigskip

with,

\bigskip

$H=\frac{\dot{R}}{R}$ \ \ \ (Hubble's parameter) \ \ \ .

\bigskip

Then we find after a short calculation:

\bigskip

$U=\sqrt{-g}\left[  6H^{2}-\frac{2}{r^{2}R^{2}}\left(  1-\frac{r^{2}}{R^{2}%
}\right)  ^{2}\right]  $ \ \ \ \ \ \ \ \ \ ,

\bigskip

and, then we find:

\bigskip

$P_{0}=\frac{4\pi^{2}}{\kappa}R(t)$ \ \ \ \ \ \ \ \ \ \ \ \ .

\bigskip

$P_{1}=P_{2}=P_{3}=0$ \ \ \ \ \ \ \ \ \ \ \ \ \ \ \ .

\bigskip

The time-varying result for \ \ $P_{0}$\ \ shows that only Cartesian
coordinates must be employed when applying pseudotensors in General
Relativity. In reference (York Jr, 1980) it is stated that, for closed
Universes, the only acceptable result is $P_{0}=0$ \ \ .

\bigskip

\textbf{3B. Flat Robertson-Walker's Counter-Example:}

\bigskip

We now repeat succinctly the \ \ $k=0$\ \ calculation, employing polar
spherical coordinates, finding the wrong result \ \ $P_{0}=\infty$\ \ .

\bigskip

Consider Einstein's pseudotensor. We shall find:

\bigskip

$U=6\sqrt{-g}H^{2}-\frac{2}{r^{2}R^{2}}$\ \ \ \ \ \ \ \ \ \ \ \ \ \ \ \ \ \ \ \ \ \ \ \ \ \ \ \ \ \ \ ;

\bigskip

and,

$P_{0}=\int\sqrt{-g}\left[  \frac{3}{\kappa R^{2}}-\frac{1}{\kappa r^{2}R^{2}%
}\right]  d^{3}x$ \ \ \ \ \ \ \ \ \ \ \ \ \ \ \ \ \ ,

\bigskip

where,

$\sqrt{-g}=R^{3}r^{2}\sin\theta$ \ \ \ \ \ \ \ \ \ \ \ \ \ \ \ \ \ \ \ \ \ \ \ \ \ \ .

\bigskip

We find then,

\bigskip

$P_{0}=\underset{r\rightarrow\infty}{\lim}\int\frac{3Rr^{2}\sin\theta}{\kappa
}\left[  1-\frac{1}{3r^{2}}\right]  d^{3}x=\underset{r\rightarrow\infty}{\lim
}\int\frac{3Rr^{2}\sin\theta}{\kappa}\left[  1-\frac{1}{3r^{2}}\right]
r^{2}\sin\theta$ $d\theta$ $d\phi$ $dr$ \ \ \ \ \ \ \ \ \ \ .

\bigskip

In the process of integration we will find:

\bigskip

$\int\limits_{\text{ \ }0}^{\text{ \ \ \ }\infty}$ $\left(  r^{4}-\frac{1}%
{3}r^{2}\right)  dr=\infty$ \ \ \ \ \ \ .

\bigskip

This shows again, that Cartesian coordinates should be employed.\ \ 

\bigskip

\textbf{3C. Counter-counter example:}

\bigskip

While we have shown that Cartesian coordinates yield acceptable results, and
spherical coordinates may lead to inconsistencies, we shall now show that
\ $LL$\ \ pseudotensor yields a correct zero result for the energy of a closed
Robertson-Walker's Universe, even if spherical coordinates are used (Berman,
1981). \ 

\bigskip

\bigskip According to Landau-Lifshitz pseudotensor, we would have:

\bigskip

$P^{\mu}=\int\left(  -g\right)  \left[  T^{\mu0}+t_{LL}^{\mu0}\right]  d^{3}x$ \ \ \ \ \ \ \ \ \ \ .

\bigskip

We apply now the superpotential:

\bigskip

$\left(  -g\right)  \left[  T^{\mu0}+t_{LL}^{\mu0}\right]  =\underset{LL}%
{U}_{\text{ \ \ \ },\text{ }\sigma}^{\mu\left[  \nu\sigma\right]  }$\ \ \ \ \ \ \ \ \ \ ,

\bigskip

where,

\bigskip

$\underset{LL}{U}_{\text{ \ \ }}^{\mu\left[  \nu\sigma\right]  }=\frac
{1}{2\kappa}\frac{\partial}{\partial x^{\lambda}}\left[  \left(  -g\right)
\left(  g^{\mu\nu}g^{\sigma\lambda}-g^{\mu\sigma}g^{\nu\sigma}\right)
\right]  $\ \ \ \ \ \ \ \ \ \ .

\bigskip

We then find successively,

\bigskip

$P^{0}=\frac{1}{2\kappa}\int\frac{\partial^{2}}{\partial x^{\sigma}\partial
x^{\lambda}}\left[  \left(  -g\right)  \left(  g^{00}g^{\sigma\lambda
}-g^{0\sigma}g^{0\lambda}\right)  \right]  d^{3}x$\ \ \ \ \ \ \ \ \ \ $=$

\bigskip

$=\frac{1}{2\kappa}\int\frac{\partial^{2}}{\partial r^{2}}\left[
-g_{22}g_{33}\right]  d^{3}x+\frac{1}{2\kappa}\int\frac{\partial^{2}}%
{\partial\theta^{2}}\left[  -g_{11}g_{33}\right]  d^{3}x$%
\ \ \ \ \ \ \ \ \ \ $=0$ \ \ \ \ \ \ \ \ \ \ \ ,

\bigskip

where we have made use of the following results:

\bigskip

$\int\limits_{\text{ \ }0}^{\text{ \ \ \ \ }\pi}\frac{\partial}{\partial
\theta^{2}}\left(  \sin^{2}\theta\right)  d\theta=2\int\limits_{\text{ \ }%
0}^{\text{ \ \ \ \ }\pi}\cos2\theta d\theta=\frac{1}{2}\left[  \sin
2\theta\right]  _{0}^{\pi}=0$ \ \ \ \ \ \ \ \ \ \ \ \ .

\bigskip

and,

\bigskip

$\left\{  \frac{d}{dr}\frac{r^{4}}{\left[  \left(  1+\frac{r^{2}}{4}\right)
^{4}\right]  }\right\}  _{0}^{\infty}=\left\{  \frac{4r^{3}}{\left(
1+\frac{r^{2}}{4}\right)  ^{4}}-\frac{2r^{5}}{\left(  1+\frac{r^{2}}%
{4}\right)  ^{5}}\right\}  _{0}^{\infty}=0$ \ \ \ \ \ \ \ \ \ \ \ .

\bigskip

Analogously we would find that the space components of the pseudomomentum are null.

\bigskip{\LARGE 4. A novel calculation}

\bigskip So many researchers \ have dealt with the present paper%
\'{}%
s subject. Why one more paper? We shall now consider, first, why the Minkowski
metric represents a null energy Universe . Of course, it is empty. But, why it
has zero-valued energy? We resort to the result of Scwarzschild%
\'{}%
s metric, (Adler et al., 1975),

\bigskip

$E=Mc%
{{}^2}%
-\frac{GM%
{{}^2}%
}{2R}$ \ \ \ \ \ \ \ \ \ \ \ \ \ \ \ \ \ \ \ \ \ \ . \ \ \ \ \ \ \ \ \ \ \ \ \ \ \ \ \ \ \ \ \ \ \ \ \ \ \ \ \ \ \ \ \ \ \ \ \ \ \ \ \ \ \ \ \ \ \ \ \ \ \ \ \ \ \ (21)

\bigskip

If \ \ \ $M=0$ \ \ , the energy is zero,too. But when we write Scwarzschild%
\'{}%
s metric, and make \ \ the mass become zero, we obtain Minkowski metric, so
that we got the zero-energy result. Any flat RW%
\'{}%
s metric, can be reparametrized as Minkowski%
\'{}%
s (Cooperstock and Faraoni,2003; Berman, 2006; 2006a).

\bigskip

Now, the energy of the Universe, can be calculated at constant time coordinate
\ $t$ \ . In particular, the result would be the same as when
\ \ $t\rightarrow\infty$ \ \ , or, even when \ \ $t\rightarrow0$ \ \ .\bigskip
Arguments for initial null energy come from Tryon(1973), and Albrow
(1973).More recently, we recall the quantum fluctuations of Alan Guth%
\'{}%
s inflationary scenario(Guth,1981). Berman(2008) gave the Machian picture of
the Universe, as being that of a zero energy . Sciama%
\'{}%
s inertia theory results also in a zero-total energy Universe(Sciama, 1953;
Berman, 2008c).

\bigskip

Consider the possible solution for RW%
\'{}%
s metric as an accelerating Universe. The scale-factor assumes a power-law , say,

\bigskip

$R=(mDt)^{1/m}$ \ \ \ \ \ \ \ \ \ \ \ \ \ \ \ \ \ \ \ \ \ \ \ \ \ \ \ \ \ \ \ \ \ \ \ \ \ \ \ \ \ \ ,\ \ \ \ \ \ \ \ \ \ \ \ \ \ \ \ \ \ \ \ \ \ \ \ \ \ \ \ \ \ \ \ \ \ \ \ \ (22)

\bigskip

where, $\ m$ $\ $, $\ \ D=$ \ constants, \ and,

\bigskip

$m=q+1>0$
\ \ \ \ \ \ \ \ \ \ \ \ \ \ \ \ \ \ \ \ \ \ \ \ \ \ \ \ \ \ \ \ \ \ \ \ \ \ \ \ \ \ \ \ \ ,
\ \ \ \ \ \ \ \ \ \ \ \ \ \ \ \ \ \ \ \ \ \ \ \ \ \ \ \ \ \ \ \ \ (23)

\bigskip where $q$ is the deceleration parameter.

For a perfect fluid energy tensor, and a perfect gas equation of state, cosmic
pressure and energy density obey the following energy-momentum conservation
law, \bigskip(Berman, 2007, 2007a),

\bigskip

$\dot{\rho}=-3H(\rho+p)$
\ \ \ \ \ \ \ \ \ \ \ \ \ \ \ \ \ \ \ \ \ \ \ \ \ \ \ \ \ \ \ \ , \ \ \ \ \ \ \ \ \ \ \ \ \ \ \ \ \ \ \ \ \ \ \ \ \ \ \ \ \ \ \ \ \ \ \ \ \ \ \ \ \ \ (24)

\bigskip

where,

\bigskip

$p=\alpha\rho$ \ \ \ \ \ \ \ \ \ \ ($\alpha=$ \ constant larger than $-1$) \ \ \ .\ \ \ \ \ \ \ \ \ \ \ \ \ \ \ \ \ \ \ \ \ \ \ \ \ \ \ \ \ \ (25)

\bigskip

On solving the differential equation, we find, for any \ $k=0$ \ \ , \ \ $1$
\ \ \ , \ \ $-1$ \ \ ,that,

\bigskip

$\rho=\rho_{0}t^{-\frac{3(1+\alpha)}{m}}$ \ \ \ \ \ \ \ \ \ \ \ \ (\ \ $\rho
_{0}=$ \ \ constant)\ \ \ \ \ \ \ \ . \ \ \ \ \ \ \ \ \ \ \ \ \ \ \ \ \ \ \ \ \ \ \ \ \ \ \ \ \ \ \ \ \ \ (26)\ 

\ \ \ \ \ \ 

When \ \ $t\rightarrow\infty$ \ \ , from (26) we see that the energy density
becomes zero, and we retrieve an "empty" Universe, or, say, again, the energy
is zero. However, this energy density is for the matter portion, but
nevertheless, as in this case, $\ \ R\rightarrow\infty$ \ \ , all masses are
infinitely far from each others, so that the gravitational inverse-square
interaction is also null. The total energy density is null, and, so, the total
energy. Notice that the energy-momentum conservation equation does not change
even if we add a cosmological constant density, because we may subtract an
equivalent amount in pressure, and \ equation (24) remains the same.

\bigskip

{\LARGE 5. Final Comments and Conclusions}

\bigskip

The zero result for the spatial components of the energy-momentum-pseudotensor
calculation, are equivalent to the choice of a center of Mass reference system
in Newtonian theory, likewise the use of comoving observers in Cosmology. It
is with this idea in mind, that we are led to the energy calculation, yielding
zero total energy, for the Universe, as an acceptable result: we are assured
that we chose the correct reference system; this is a response to the
criticism made by some scientists which argue that pseudotensor calculations
depend on the reference system, and thus, those calculations are devoid of
physical meaning.

\bigskip

\bigskip The counter-example ( \ $k=+1$\ ) \ \ shows, nevertheless, that
Cartesian coordinates need to be used. Next, a new counter-example ( $k=0$ \ )
shows the same problem. In the following calculation, we found a
counter-counter-example, where the use of spherical coordinates, although
tragic earlier, does no harm in the Landau-Lifshitz calculation. We thank
J.Katz, for several advises, in order to allow any kind of coordinates in
energy calculations, and that superpotentials should be preferred, because our
calculations would be simplified (Katz, 2006, 1985; and Ori, 1990; et al, 1997).

\bigskip

The zero-total-energy of the Universe has been made clear. Related conclusions
by Berman should be consulted (Berman,2006,2006a,2007,2007a,b,c,d,2008,a,b).
As a bonus, we can assure that there was not an initial infinite energy
density singularity, because attached to the zero-total energy conjecture,
there is a zero-total energy-density result, as was pointed by Berman
elsewhere (Berman, 2008).The so-called total energy density of the Universe,
which appears in some textbooks, corresponds only to the non-gravitational
portion, and the zero-total energy density results when we subtract from the
former, the opposite potential energy density.

\bigskip As Berman(2009) shows, we may say that the Universe is
\emph{singularity -free}, and was created \emph{ab-nihilo}.

\bigskip In order to close forever this subject, some words follow. The
equivalence principle, says that at any location, spacetime is \ (locally)
flat, and a geodesic coordinate system may be constructed, \textit{where the
Christoffel symbols are null. The pseudotensors are, then, at each point,
null. But now remember that our old Cosmology requires a co-moving observer at
each point}. \textit{It is this co-motion that is associated with the geodesic
system, and, as RW%
\'{}%
s metric is homogeneous and isotropic, for the co-moving observer, the
zero-total energy density result, is repeated from point to point, all over
spacetime. Cartesian coordinates are needed, too, because curvilinear
coordinates are associated with fictitious or inertial forces, which would
introduce inexistent accelerations that can be mistaken \ additional
gravitational fields (i.e.,that add to the real energy). Choosing Cartesian
coordinates is not analogous to the use of center of mass \ \ frame in
Newtonian theory, but the null results for the spatial components of the
pseudo-quadrimomentum show compatibility. }

\bigskip

Berman (2009) has discussed why there is no zero-time infinite energy-density singularity.

\textit{\bigskip}The calculation of Section 4, is original, too.

\bigskip

{\LARGE Acknowledgements}

\bigskip

I thank the opportunity given by the present referee, who allowed me to revise
and enlarge the manuscript more efficiently, by following his valuable report.

The author thanks Marcelo Fermann Guimar\~{a}es, Nelson Suga, Mauro Tonasse,
Antonio F. da F. Teixeira, and for the encouragement by Albert, Paula and
Geni. The advisor for my M.Sc. Thesis (Berman,1981), was, my present friend
and colleague, Prof. Fernando de Mello Gomide, a full-fledged scientist, and
the "father" of Cosmology in Brazil.

\bigskip

\bigskip{\Large References}

\bigskip Adler, R.; Bazin, M,; Schiffer, M. (1975) - \textit{Introduction to
General Relativity}\ - 2nd. edtn., McGraw-Hill, N.Y.

Albrow,M.G.(1973) - Nature, \textbf{241},56.

Banerjee, N.; Sen, S. (1997) - Pramana J.Phys., \textbf{49}, 609.

Berman, M. S. (1981, unpublished) - M.Sc. thesis, Instituto Tecnol\'{o}gico de
Aeron\'{a}utica, S\~{a}o Jos\'{e} dos Campos, Brazil.Available online, through
the federal government site \ www.sophia.bibl.ita.br/biblioteca/index.html
(supply author%
\'{}%
s surname and keyword may be "pseudotensor"or "Einstein").

Berman, M. S. (2006) - Chapter 5 of: \textit{Trends in Black Hole Research},
ed by Paul V. Kreitler, Nova Science, New York.

Berman, M. S. (2006a) - Chapter 5 of: \textit{New Developments in Black Hole
Research}, ed by Paul V. Kreitler, Nova Science, New York.

Berman, M. S. (2007) - \textit{Introduction to General Relativity and the
Cosmological Constant Problem, }Nova Science, New York.

Berman, M. S. (2007a) - \textit{Introduction to General Relativistic and
Scalar Tensor}

\textit{Cosmologies }, Nova Science, New York.

Berman, M. S. (2007b) - Astrophysics and Space Science, \textbf{312}, 275.

Berman, M. S. (2007c) - RevMexAA, \textbf{43,} 297-301.

Berman, M. S. (2007d) - Astrophysics and Space Science, \textbf{311}, 359.

Berman, M. S. (2008) - \textit{A Primer in Black Holes, Mach's Principle and
Gravitational Energy,} Nova Science Publishers, New York.

Berman, M. S. (2008a) - \textit{A General Relativistic Rotating Evolutionary
Universe, }Astrophysics and Space Science, \textbf{314, }319-321.

Berman, M. S. (2008b) - \textit{A General Relativistic Rotating Evolutionary
Universe - Part II, }Astrophysics and Space Science, \textbf{315, }367-369.

Berman, M. S. (2009) - \textit{General Relativistic Singularity-free
Cosmological Model, }Astrophysics and Space Science, \textbf{321}, 157-160.

\bigskip Carmeli,M.;Leibowitz,E.;Nissani,N.(1990)-\textit{Gravitation:SL(2,C)
Gauge Theory and Conservation Laws,}World Scientific, Singapore.

Cooperstock, F.I. (1994) - GRG \textbf{26}, 323.

Cooperstock, F.I.; Israelit,M. (1995) - \textit{Foundations of Physics},
\textbf{25}, 631.

\bigskip Cooperstock, F.I.;Faraoni,V.(2003) - Ap.J. 587,483.

Feng, S.; Duan, Y. (1996) - Chin. Phys. Letters, \textbf{13}, 409.

Feynman, R. P. (1962-3) - \textit{Lectures on Gravitation} , Addison-Wesley, Reading.

Freud, P.H.(1939) - Ann. Math, \textbf{40}, 417.

Garecki, J.\ (1995) - GRG, \textbf{27}, 55.

Guth, A. (1981) - Phys. Rev. \textbf{D23}, 347 .

Hawking, S. (2001) - \textit{The Universe in a Nutshell, }Bantam, New York.

Johri, V.B.; et al. (1995) - GRG, \textbf{27}, 313.

Katz, J. (1985) - Classical and Quantum Gravity \textbf{2}, 423.

Katz, J. (2006) - Private communication.

Katz, J.; Ori, A. (1990) - Classical and Quantum Gravity\textbf{ 7}, 787.

Katz, J.; Bicak, J.; Lynden-Bell, D. (1997) - Physical Review \textbf{D55}, 5957.

Landau, L.; Lifshitz, E. (1975) - \textit{The Classical Theory of Fields},
4th. Revised ed.; Pergamon, Oxford.

Radinschi, I. (1999) - Acta Phys. Slov., \textbf{49}, 789. Los Alamos
Archives, gr-qc/0008034.

Rosen, N.(1994) - \textit{Gen. Rel. and Grav.} \textbf{26}, 319.

Rosen, N.(1995) - GRG, \textbf{27}, 313.

So, L.L.; Vargas, T. (2006) - Los Alamos Archives, gr-qc/0611012 .

Tryon, E.P.(1973) - Nature, \textbf{246}, 396.

Weinberg, S. (1972) - \textit{Gravitation and Cosmology} , Wiley, New York.

Xulu, S. (2000) - International Jounal of Theoretical Physics, \textbf{39},
1153. Los Alamos Archives, gr-qc/9910015 .

York Jr, J.W. (1980) - \textit{Energy and Momentum of the Gravitational
Field}, in \textit{A Festschrift for Abraham Taub}, ed. by F.J. Tipler,
Academic Press, N.Y.

\end{document}